\documentclass[conference] {IEEEtran}

\IEEEoverridecommandlockouts

\usepackage{amsmath,amssymb,amsfonts}
\usepackage{algorithmic}
\usepackage{graphicx}
\usepackage{textcomp}
\usepackage{xcolor}
\usepackage[fontsize=11pt]{fontsize}

\usepackage{tikz}
\usepackage[american,cuteinductors,smartlabels,nooldvoltagedirection]{circuitikz}
\usetikzlibrary{backgrounds}
\usepackage{siunitx}

\usepackage{hyperref} 

\usepackage{amsthm}
\newtheorem{ass}{Assumption}
\newtheorem*{cond}{Condition}

\theoremstyle{remark}
\newtheorem*{remark}{Remark}
\newtheorem{condrem}{Condition Remark}

\usepackage[caption=false]{subfig}
\usepackage[sorting=none, style=ieee, urldate =comp]{biblatex} 
\usepackage{booktabs}
\newcommand{\addpic}[1]{\includegraphics[width=.23\textwidth]{#1}}
\newcommand{\addpec}[1]{\includegraphics[width=.475\textwidth]{#1}}
\newcolumntype{C}{>{\centering\arraybackslash}m{.23\textwidth}}
\newcolumntype{E}{>{\centering\arraybackslash}m{.475\textwidth}}

\usepackage{csquotes}

\def\BibTeX{{\rm B\kern-.05em{\sc i\kern-.025em b}\kern-.08em
    T\kern-.1667em\lower.7ex\hbox{E}\kern-.125emX}}

\begin{document}

    \title{

    \LARGE
    Analysis of potential lifetime extension through dynamic battery reconfiguration\\
    
    \vspace{0.5em}
    
    \begin{center}
    {
        \large Albert~Škegro, 
        Changfu~Zou,~\IEEEmembership{\large Senior Member,~IEEE}, 
        Torsten~Wik,~\IEEEmembership{\large Member,~IEEE}
        
        CHALMERS UNIVERSITY OF TECHNOLOGY 
        
        412 96 Göteborg, Sweden
        
        Tel.: +46 / (31) 772 10 00 
        
        E-Mail: \{albert.skegro, changfu.zou, torsten.wik\}@chalmers.se  
        
        URL: https://www.chalmers.se/       
        }      

    \end{center}
    
    \vspace{-2em}
}

\maketitle

    \section*{Acknowledgment}
    This work was funded by Mistra Innovation 23 under the project BattVolt. The computations were enabled by resources provided by Chalmers e-Commons at Chalmers.

    \vspace{0.4cm} 
    \begin{IEEEkeywords}
        Batteries, Battery management systems (BMS), Modular reconfigurable batteries, Lifetime, Switching cells.
    \end{IEEEkeywords}

    \vspace{0.4cm}         
    \begin{abstract}
        Growing demands for electrification result in increasingly larger battery packs. Due to factors such as cell position in the pack and variations in the manufacturing process, the packs exhibit variations in the performance of their constituent cells. Moreover, due to the fixed cell configuration, the weakest cell renders the pack highly susceptible to these variations. Reconfigurable battery pack systems, which have increased control flexibility due to additional power electronics, present a promising solution for these issues. Nevertheless, to what extent they can prolong the battery lifetime has not been investigated.
        
        This simulation study analyzes the potential of dynamic reconfiguration for extending battery lifetime w.r.t. several parameters. Results indicate that the lifetime extension is larger for series than for parallel configurations. For the latter, the dominant factor is equivalent full cycles spread at the end of life, but resistance increase with age and the number of cells in parallel are also influential. Finally, for the former, the number of series-connected elements amplifies these effects.
    \end{abstract}
    
    \vspace{0.4cm} 
    \section{Introduction}\label{sec:Introduction}

    Contemporary transportation systems rely significantly on fossil fuels, whose consumption and associated tailpipe emissions not only damage ecosystems but are also main contributors to climate change and are unsustainable in the long run. 
    
    Electrification based on lithium-ion (Li-ion) batteries presents a promising way to mitigate this dependency. In order to attain their specifications, battery pack systems for electric vehicles (EV) are often comprised of many Li-ion battery cells. For instance, Tesla Model S Plaid has nearly 8000 cells~\cite{Kane:12}, and a fully electric truck produced by Scania might have more than 20000 cells~\cite{Sve:20}. The battery management system (BMS), which accurately monitors the cells in the pack and keeps them within a safe (electric-thermal) operating window, is an indispensable component of such a large battery pack system. 

    In spite of this, large battery pack systems tend to be used inefficiently. As a result of variations in the manufacturing process and different local conditions in the pack, the cells do not age uniformly. A consequent large spread in cell capacities, or cell internal resistances, results in significant performance degradation for the fixed cell configuration with regard to capacity utilization and power output. Furthermore, the inability to disconnect cells implies that failure of a single cell not only presents a safety hazard but also renders the remaining healthy cells in the pack unusable. 

    Dynamic battery reconfiguration is a promising concept in this regard~\cite{9300283}. Essentially, battery pack systems allowing for dynamic reconfiguration, i.e., reconfigurable battery pack systems (RBSs), enable greater flexibility during the battery pack operation than conventional battery pack systems with a fixed cell configuration (FBSs). The reconfiguration is achieved by employing additional power electronics, e.g., by placing a certain number of switches around each cell, or group of cells, thus allowing for finer cell-level control and a range of potential benefits such as enhanced fault tolerance, prolonged battery pack life, customized output voltage, and mixing of cells having different properties. Promising RBS applications include stationary energy storage systems (e.g., DC charging and microgrids~\cite{Engelhardt}), as well as battery pack systems for electric vehicles (e.g.,~\cite{Bouchhima:17},~\cite{Huang:21}). Nevertheless, RBSs are not without challenges of their own. Although the additional power electronics enhances the operating flexibility, it also results in further electric and thermal losses. Moreover, the increased number of components also yields a higher total cost. Finally, increasing the number of switches for the purpose of improved controllability may render the system management highly complex.

    The field of dynamic battery reconfiguration has been approached in several survey articles which have focused on the overview of the concept of dynamic battery reconfiguration~\cite{9300283}, the comparison of existing RBS hardware configurations~\cite{Muhammad:19}, and the review of control strategies in existing RBSs~\cite{Pinter:21}. 

    A systematic analysis of improvements gained by utilizing RBSs in place of FBSs is not only important for the motivation of RBSs but also presents a good initial step in designing an RBS solution. 
    For instance, to what extent dynamic battery reconfiguration should be implemented will strongly depend on the lifetime extension of an RBS relative to the variations in its cells. No such analysis has been presented to the best of the authors' knowledge.
    Therefore, this paper aims to fill this gap by providing a simulation-based analysis with the following main contributions: 
    \begin{itemize}
        \item A systematic method to create a set of cell ageing models from provided cell ageing data as an input to the
        study of the lifetime extension.
        \item A method for simulation-based analysis of the potential lifetime extension.
        \item Formulation of the lifetime extension for different RBS configurations.
    \end{itemize}

    The paper is organized as follows. Section~\ref{sec:Method} outlines the method used for the analysis, including the necessary definitions and the underlying assumptions. Section~\ref{sec:Implementation} presents the implementation. Section~\ref{sec:Results} discusses the results for a publicly available dataset. Finally, Section~\ref{sec:Conclusion} summarizes the findings. 

    \section{Method}\label{sec:Method}

    \subsection{The system}

        In this paper, the term \emph{parallel cell unit} (PU), denoting an electrical unit comprising \(N_p\) parallel-connected cells, is introduced. The PU can refer to either the \emph{PU without reconfiguration}, i.e., the PU with a fixed cell configuration (FPU) or the \emph{PU with ideal reconfiguration}, i.e., the reconfigurable PU (RPU). 
        In the RPU, there is a switch adjacent to every cell which can engage or disengage the cell. Furthermore, each RPU is connected to a pair of switches, ensuring that the RPU can be engaged or bypassed. 
        
        Moreover, the term \emph{generalized module} (GM) is used to denote an electrical unit comprising \(N_s\) series-connected PUs. The GM can refer to either the \emph{GM without reconfiguration}, i.e., the GM with a fixed cell configuration (FGM) consisting of \(N_s\) series-connected FPUs or the \emph{GM with ideal reconfiguration}, i.e., the reconfigurable GM (RGM) consisting of \(N_s\) series-connected RPUs.
    
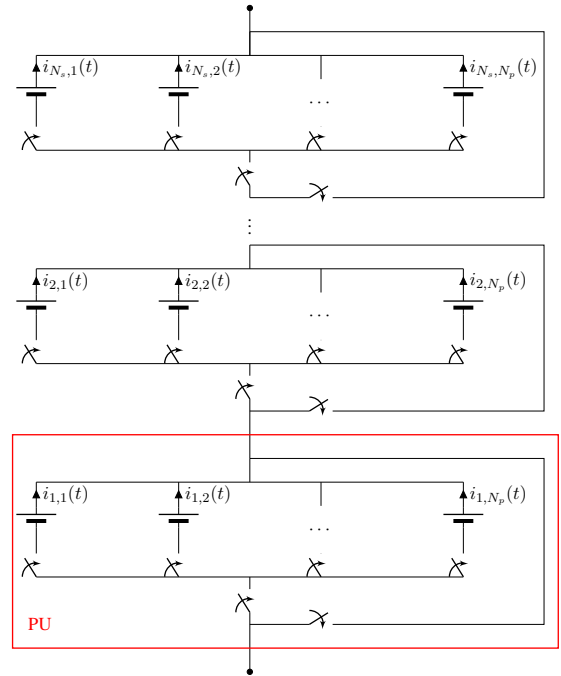
\begin{figure}
            \setlength\belowcaptionskip{-0.9\baselineskip}        
            \begin{center}
                \resizebox{0.85\columnwidth}{!}
                {
                \centering
                \begin{tikzpicture}
    \begin{circuitikz}
                
        \foreach \i in {0,1,2} 
        {
            \ifthenelse{\i = 0}
                {\draw (5,4.5*\i) 
                    node[circ] at (5,4.5*\i) {};
                }
                {}
            
            \ifthenelse{\i = 2}
                {
                \draw (5,4.5*\i+0.5)                    
                    node {$\vdots$};
                \draw (5,4.5*\i + 1) 
                    to[spst] ++(0,1);                       
                }
                {
                \draw (5,4.5*\i)                    
                    to[short] ++(0,1)
                    to[spst] ++(0,1);                    
                }

            \draw (5,4.5*\i+1)
                to[short] ++(1,0)
                to[spst] ++(1,0)
                to[short] ++(4.2,0)
                to[short] ++(0,3.5)
                to[short] ++(-6.2,0)
                node (A4) {};  

            \draw (9.5,4.5*\i+2)                
                to[short] (0.5,4.5*\i+2);                            

            \draw (0.5,4.5*\i+2) 
                to[spst] ++(0,0.5);         
            \draw (3.5,4.5*\i+2) 
                to[spst] ++(0,0.5);    
            \draw (6.5,4.5*\i+2) 
                to[spst] ++(0,0.5);    
            \draw (9.5,4.5*\i+2) 
                to[spst] ++(0,0.5);   

            \ifthenelse{\i = 2}
                {        
                \draw (0.5,4.5*\i+4)        
                    to [battery2,i=$i_{N_s,1}(t)$] ++(0,-1.5);  
                \draw (3.5,4.5*\i+4)
                     to [battery2,i=$i_{N_s,2}(t)$] ++(0,-1.5); 
                \draw (6.5,4.5*\i+4) 
                    to[short] ++(0,-0.5);
                \draw (6.5,4.5*\i+3)
                    node {$\cdots$}; 
                \draw (9.5,4.5*\i+4)                
                     to [battery2, i=$i_{N_s,N_p}(t)$] ++(0,-1.5); 
                    }
                {
                \ifthenelse{\i = 1}
                {        
                \draw (0.5,4.5*\i+4)        
                    to [battery2,i=$i_{2,1}(t)$] ++(0,-1.5);  
                \draw (3.5,4.5*\i+4)
                     to [battery2,i=$i_{2,2}(t)$] ++(0,-1.5); 
                \draw (6.5,4.5*\i+4) 
                    to[short] ++(0,-0.5);
                \draw (6.5,4.5*\i+3)
                    node {$\cdots$}; 
                \draw (9.5,4.5*\i+4)                
                     to [battery2, i=$i_{2,N_p}(t)$] ++(0,-1.5); 
                    }
                {        
                \draw (0.5,4.5*\i+4)        
                    to [battery2,i=$i_{1,1}(t)$] ++(0,-1.5);  
                \draw (3.5,4.5*\i+4)
                     to [battery2,i=$i_{1,2}(t)$] ++(0,-1.5); 
                \draw (6.5,4.5*\i+4) 
                    to[short] ++(0,-0.5);
                \draw (6.5,4.5*\i+3)
                    node {$\cdots$}; 
                \draw (9.5,4.5*\i+4)                
                     to [battery2, i=$i_{1,N_p}(t)$] ++(0,-1.5); 
                    }
                     }

            \draw (9.5,4.5*\i+4) 
                to[short] (0.5,4.5*\i+4);   

            \ifthenelse{\i = 1}
            {\draw (5,4.5*\i+4)  
                to[short] ++(0,0.5);            
                }
            {\draw (5,4.5*\i+4)  
                to[short] ++(0,0.5);
                }
            \ifthenelse{\i = 2}
            {\draw (5,4.5*\i + 4) 
                to[short] ++(0,0.5)   
                to[short] ++(0,0.5)   
                node[circ] at (5,4.5*\i + 5) {};}
            {}   
                
        }
        
            \draw[red, thick] (0,0.5) rectangle (11.5,5)
                node[pos=0.05, above]{PU};                    

    \end{circuitikz}
\end{tikzpicture}
                }
                \caption{A reconfigurable generalized module (RGM)
                with one of the reconfigurable parallel cell units (RPUs) enclosed in red, $N_p$ representing the number of cells in the RPU, and $N_s$ denoting the number of series-connected RPUs in the RGM.}
                \label{fig:GRU}
            \end{center}
        \end{figure}

        RGM and its fundament RPU, depicted in Fig.~\ref{fig:GRU}, form the basis of the subsequent analysis.

    \subsection{Assumptions and conditions}

        In order to allow for the quantitative analysis, a considerable amount of full lifetime simulations for many cells is required. For this to be feasible, a set of assumptions is formulated to appropriately limit the scope of the study, or to simplify the simulation process without loss of generality.
         
        \begin{ass}\label{Ass:PDFs_BOL_EOL}
            The probability distribution functions (PDFs) of cell capacities at the beginning of life (BOL), denoted as \({Q}_{c,s}\), and of cell equivalent full cycles (EFCs) at the end of life (EOL), denoted as \(EFC_{c,e}\), are known.
        \end{ass}

        \begin{remark}
            This assumption is the cornerstone of the subsequent analysis. Within the following simulation-based analysis, normal distribution with mean \(\mu_{s}\) and variance \(\sigma^{2}_{s}\) is chosen for the PDF of cell capacities at the BOL in line with experimental results obtained for a statistically significant number of Li-ion battery cells~\cite{PAUL2013642}. 
            Hence, for the cell capacities at the BOL, normalised by the cell nominal capacity \(Q_{c,nom}\), we have            \begin{equation}\label{eq:Q_distribution}
                \widetilde{Q}_{c,s} \sim \mathcal{N}\left(\mu_{s},\,\sigma^{2}_{s}\right)\, , 
            \end{equation} 
            where \(\widetilde{Q}_{c,s}(t) = Q_{c,s}(t)/Q_{c,nom}\).    
            
            Furthermore, considering the PDF of cell EFCs at the EOL, normal distribution with mean \(\mu_{e}\) and variance \(\sigma^{2}_{e}\) is selected to keep the simulations comparable and simple, i.e.  
            \begin{equation}\label{eq:EFC_distribution}
                EFC_{c,e} \sim \mathcal{N}(\mu_{e},\,\sigma^{2}_{e})\, .
            \end{equation}               
             However, any given distribution can equally be applied.
        \end{remark}

        \begin{ass}\label{Ass:ThermalDyn}
            The thermal dynamics does not affect the  
            lifetime difference between RBSs and FBSs.
        \end{ass}
   
        \begin{remark}
           The cell temperature rise is expected to be small for the operating conditions. Hence, the cell temperatures are fixed and are assigned a value of \(25 ^{\circ}\)C, considering that it is a typically desired operating temperature for Li-ion battery cells. 
        \end{remark}

        \begin{ass}\label{Ass:Cycleageing}
            The cell capacity degradation is dominated by cycling ageing.
        \end{ass}

        \begin{remark}
             In order to assess cell ageing, cycling tests are often used. 
             Based on a publicly available experimental cell ageing dataset from~\cite{Li:21-1}, within this analysis, it is assumed that the cells are in operation during most of their lives and that the difference in calendar ageing between RBS and FBS is negligible.
        \end{remark}

        \begin{ass}\label{Ass:CapFadeLinear}
            The cell capacity fade can be modelled as a linear function of cell EFC.
        \end{ass}

\begin{figure}
            \begin{center}
                \includegraphics[width=\linewidth]{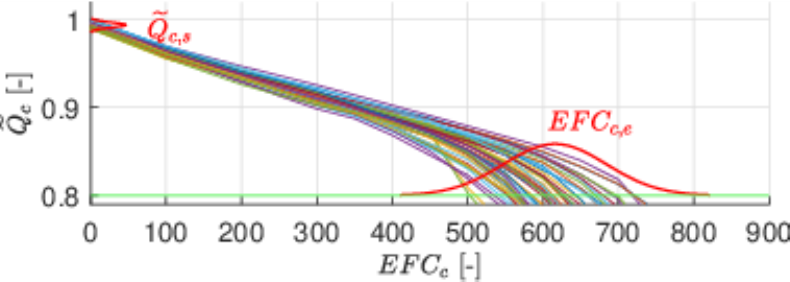}  
                \caption{Trajectories of Li-ion battery cell capacity fade with an illustration of the PDFs of normalised cell capacities at the BOL and of cell EFC at the EOL (in red). The data originate from~\cite{Li:21-1}.
                }
                \label{fig:ageingModel_dQ}
            \end{center}
        \end{figure}

        \begin{remark}
            In the initial stage of cell ageing, the cell capacity fade process can often be well approximated using a first-order polynomial~\cite{Preger_2020}. Nevertheless, the cell undergoes different ageing processes during its lifetime, and accelerated and nonlinear ageing may appear when it approaches its EOL~\cite{zhang2022machine}. 
            While leaving the study of different types of ageing models for future research, within this analysis, linear cell capacity fade trajectories, constructed based on cell ageing dataset~\cite{Li:21-1}, are considered.
            The data are shown in Fig.~\ref{fig:ageingModel_dQ}.
        \end{remark}

\begin{figure}
            \begin{center}
                \includegraphics[width=\linewidth]{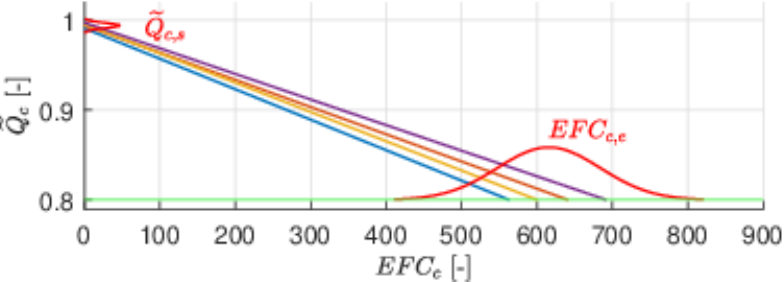} 
                \caption{Illustrative example of cell capacity fade trajectories for a PU consisting of four cells.}
                \label{fig:ageingModels}
            \end{center}
        \end{figure}

        Consequently, the normalised cell capacity fade is modelled by connecting one realization of the distribution (\ref{eq:Q_distribution}) with one realization of the distribution (\ref{eq:EFC_distribution}), as illustrated in Fig.~\ref{fig:ageingModels}.

        Denormalisation by \(Q_{c,nom}\) then results in        \begin{equation}\label{eq:EFCfromQ_EOL}
            EFC_{c}(t) = a_{c} - b_{c}Q_{c}(t)  \, ,
        \end{equation}     
        where \(a_{c} > 0\), \(b_{c} > 0\), and \(\widetilde{Q}_{c}(t) = Q_{c}(t)/Q_{c,nom}\).

        \begin{ass}\label{Ass:ResIncLinear}
            The cell resistance increase is proportional to the cell capacity fade.
        \end{ass}
        
\begin{figure}
            \begin{center}
                \includegraphics[width=\linewidth]{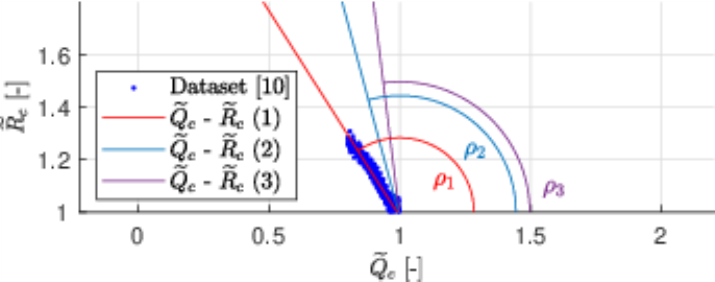}
                \caption{The relationship between the normalised cell capacity fade and the normalised cell resistance increase (in dots for experimental data from \cite{Li:21-1}), along with the line of best fit for experimental data (1) and two perturbations for the simulation-based analysis ((2), (3)).}
                \label{fig:ageingModel_dQ_dR}
            \end{center}
        \end{figure}

        \begin{remark}
            This assumption builds upon
            experimental results \cite{PASTORFERNANDEZ2016574}. A representative depiction of the relationship between the normalised cell capacity fade and the normalised cell resistance increase can be found in Fig.~\ref{fig:ageingModel_dQ_dR} where the normalised cell resistance $\widetilde{R}_{c}$ is defined as $\widetilde{R}_{c}(t)  = R_{c}(t)/R_{c,nom}$.
        \end{remark}

        Hence,
        \(\widetilde{Q}_{c}\) and 
        \(\widetilde{R}_{c}\) are linked by the relationship
        \begin{equation}\label{eq:PropFactor_b}        
            \widetilde{R}_{c}(t) = -k_{RQ}\widetilde{Q}_{c}(t) + l_{RQ}  \, ,
        \end{equation}        
        where \(k_{RQ} > 0\), and \(l_{RQ} > 0\).
        
        Considering (\ref{eq:PropFactor_b}), the angle \(\rho\) between \(\widetilde{R}_{c} = 1\) and the \(\widetilde{Q}_{c}-\widetilde{R}_{c}\) lines from Fig.~\ref{fig:ageingModel_dQ_dR} can be calculated as
        \begin{equation}\label{eq:rho}
            \rho_m = \arctan \left(1-l_{RQ,m} \right)\,, \quad m = {1,2,3} \, .
        \end{equation}        

        Denormalising (\ref{eq:PropFactor_b}) and combining it with
        (\ref{eq:EFCfromQ_EOL}) results in
        \begin{equation}\label{eq:ResIncrease_tot}
            \begin{split}            
                R_{c}(t) = c_{c} + d_{c} EFC_{c}(t)  \, ,            
            \end{split}
        \end{equation}               
        where \(c_{c} > 0, \, d_{c} > 0\).

        \begin{ass}\label{Ass:OCVSOCfixed}
            The relationship between the open-circuit voltage (OCV) and the state-of-charge (SOC), namely the OCV-SOC curve, changes insignificantly during cell ageing.
        \end{ass}

        \begin{remark}
            Within this analysis, the OCV-SOC relationship is, for the sake of simplicity, assumed to be constant, in line with the work in \cite{SONG2021100091}.  
        \end{remark}

        \begin{ass}\label{Ass:IdealRec}
            Full reconfiguration can be used to control cell usage such that all cells have the same capacity and SOC at the EOL.
        \end{ass}

        \begin{remark}
            Consequently, when RBS reaches its EOL, its cells are at the same lower voltage limit \(v_{min}\).
        \end{remark}

        \begin{ass}
            Losses in the RBS switches are negligible.
        \end{ass}

        \begin{remark}
            Losses in the switches consist of switching losses and conduction losses. While switching losses are relatively low due to a low switching frequency commonly designed for RBSs~\cite{Engelhardt}, 
            results from~\cite{DoubleString} 
            indicate that conduction losses tend to contribute by less than 2\% to the total RBS losses. 
        \end{remark}

        \begin{cond}
            Repeated cycles of discharging using constant current (CC) of one C-rate (1C) until the lower voltage limit is reached and charging using the constant current--constant voltage (CC--CV) protocol are applied. CC charging using 1C is executed until the upper voltage limit, followed by the CV phase until the charging current has decreased to C/30. 
        \end{cond}

        \begin{condrem}\label{condrem:CondRem1}
            This simulation setup is in line with (even lower) C-rates expected in most EV operations (cf.  \cite{CRateHist}). 
        \end{condrem}
        \begin{condrem}\label{condrem:CondRem2}
            Due to CC discharging and CC-CV charging, the polarization voltage transients are negligible in duration compared to both the discharge and the charge phase of the repeated cycles. Thus, the cell electric model reduces to a zero-order equivalent circuit model (ECM).        
        \end{condrem}

        \begin{figure*}
          \centering
            \subfloat{\includegraphics[width=.47\textwidth]{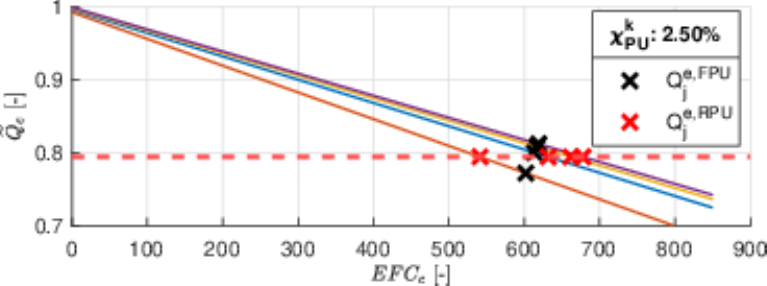}}\quad
            \subfloat{\includegraphics[width=.47\textwidth]{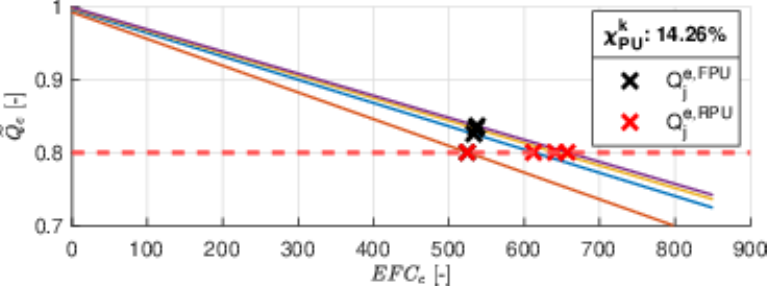}}\quad
            \caption{Comparison of cell capacities at the EOL of FPU and RPU, each with four cells, evaluated for Approach 1 (left) and 2 (right).}
            \label{fig:EOL_PU}
        \end{figure*}

    \subsection{Definitions}    
        Two approaches are used to define the EOL of PU. Approach 1 is based on commonly used reference performance testing procedures (e.g. \cite{groot2012state}) and relates to usable capacity in EV applications. Approach 2 is focused on cell safety instead, implying that if the capacity of any cell in the PU is reduced below a certain threshold, it poses a safety hazard (cf. \cite{Remus}). The latter approach naturally assumes knowledge of all cell capacities in the PU. 
        
        In subsequent definitions, index \enquote{\(i\)} refers to PU \(i\) in the GM \(\left(i=1, \ldots, N_s\right)\), while index \enquote{\(j\)} refers to cell \(j\) in the PU \(\left(j=1, \ldots, N_p\right)\). The negative discharge current sign convention applies throughout the calculations.

        \emph{The PU 1C-discharge current} is conveniently defined based on the cell 1C-discharge current:
        \begin{equation}\label{eq:1C_PU}
            i_{PU}^{1C} = i_{c}^{1C} \, N_p \, .
        \end{equation}        

        \emph{The PU nominal 1C-capacity} is determined as the ampere-hour quantity obtained by cycling the FPU in the first CC discharge cycle with \(i_{PU}^{1C}\) from the higher voltage level \(v_{max}\) (corresponding to \(100 \%\) SOC, starting at \(t_{dis,1}^{s}\)), to the lower voltage level \(v_{min}\) (reached at \(t_{dis,1}^{e}\)), in line with \cite{reuter2019importance} and \cite{team2008guide}:
        \begin{equation}\label{eq:Q_PU_nom}
            Q_{PU,nom}^{1C} = -\int_{t_{dis,1}^{s}}^{t_{dis,1}^{e}}i_{PU}^{1C}\,d\tau \, . 
        \end{equation}       

        The capacity trajectories of all PUs are assumed to be measured every cycle (hence, they are piecewise constant) and are calculated analogously to (\ref{eq:Q_PU_nom}). For discharge cycles \(w\) (starting at \(t_{dis,w}^{s}\) and finishing at \(t_{dis,w}^{e}\)) and \(w+1\) (finishing at \(t_{dis,w+1}^{e}\)), i.e., for \{\(t : t_{dis,w}^{e} < t < t_{dis,w+1}^{e} \)\}, the \emph{PU 1C-capacity} is defined as
        \begin{equation}\label{eq:Q_PU_1C}
            \begin{split}
                Q_{PU}^{1C}(t) 
                &= -\int_{t_{dis,w}^{s}}^{t_{dis,w}^{e}}i_{PU}^{1C}\,d\tau \, .
            \end{split}
        \end{equation}            

        In Approach 1, the \emph{EOL of PU} is defined as the time instant when the PU 1C-capacity from (\ref{eq:Q_PU_1C}) reaches \(80\,\%\) of the PU nominal 1C-capacity, defined in (\ref{eq:Q_PU_nom}):
        \begin{equation}\label{eq:PU_EOL_Method1}
            t_{PU}^{e} = \min_t \, \, s.t. \, \,   \,Q_{PU}^{1C}(t) \leq 0.8\,  Q_{PU,nom}^{1C} \, .
        \end{equation}  

        In Approach 2, the \emph{EOL of PU} is defined as the time instant when the first of the cells in the PU reaches the threshold of \(80\,\%\) of the cell nominal capacity: 
        \begin{equation}\label{eq:PU_EOL_Method2}
            t_{PU}^{e} = \max_t \, \, s.t. \, \,   \,Q_{j}(t) \geq 0.8 \, Q_{c,nom}\,,\quad \forall j \, .
        \end{equation}

    \subsection{Analysis}   

        The evaluation of potential lifetime extension through dynamic battery reconfiguration entails two steps. The first step necessitates quantifying the lifetime extension by employing RPU instead of FPU. This quantification is accomplished through a comparison of the cell EFCs at the EOL of FPU and RPU. These EFCs are calculated based on the respective cell capacities at the EOL using (\ref{eq:EFCfromQ_EOL}). This section therefore details how the cell capacities at the EOL of FPU and RPU are determined.

        Subsequently, the second step quantifies the lifetime extension by utilizing RGM in place of FGM. As RPU is the fundament of RGM, results obtained in the first step provide the necessary input for this quantification.          
        
        \subsubsection{PU analysis}\label{subsubsec:Analysis_PUcase} 

            For FPU, the full lifetime simulation is needed to determine the cell capacities at its EOL. For RPU, though, the cell capacities at its EOL can be calculated starting from the Assumptions. Hence, the idea is to repeatedly pairwise pick realizations of cell capacity at the BOL and cell EFC at the EOL (representatively depicted in Fig.~\ref{fig:ageingModels}); to determine cell capacities at the EOL of FPU from the full lifetime simulation; to determine cell capacities at the EOL of RPU through calculation (for both Approach 1 and 2); and finally, to evaluate the lifetime extension through dynamic battery reconfiguration, with an illustrative end-result depicted in Fig.~\ref{fig:EOL_PU}.

            \paragraph{FPU performance}\label{par:PU_without_rec}
                The EOL of FPU is reached at \(t_{FPU}^{e}\). The cell capacities at \(t_{FPU}^{e}\) are obtained from the full lifetime simulation:
                \begin{equation}\label{eq:CellCap_EO_PU_withoutrec}
                    Q_{j}^{e,FPU} \mathrel{\widehat{=}} Q_{j}\left(t_{FPU}^{e}\right)\,,\quad \forall j \, .
                \end{equation}             
                
                Subsequently, the EFC of FPU at \(t_{FPU}^{e}\) is obtained by summing the cell EFCs at \(t_{FPU}^{e}\) and using~(\ref{eq:EFCfromQ_EOL}):
                \begin{equation}\label{eq:EFC_EOL_noreconf}
                    \begin{aligned}
                        EFC_{FPU}^{e} 
                        &= \sum_{j=1}^{N_p}EFC_{j}^{e,FPU}\\
                        &= \sum_{j=1}^{N_p} \left(a_{j}-b_{j}Q_{j}^{e,FPU}\right)                 
                    \end{aligned}
                \end{equation}

            \paragraph{RPU performance}\label{par:PU_with_rec}
                The EOL of RPU is reached at \(t_{RPU}^{e}\). Assumption~\ref{Ass:IdealRec} indicates the equivalence of cell capacities and cell SOCs at \(t_{RPU}^{e}\). The latter implies the equivalence of cell OCVs at \(t_{RPU}^{e}\), i.e.:
                \begin{subequations}\label{eq:optim}
                    \begin{align}
                        Q_{j}^{e,RPU}
                        &\mathrel{\widehat{=}}& Q_{j}\left(t_{RPU}^{e}\right)  &\mathrel{\widehat{=}}& Q^{e,RPU} \,,  \enspace \forall j \, ,\label{eq:cost}\\    z_{j}^{e,RPU}&\mathrel{\widehat{=}}&z_j\left(t_{RPU}^{e}\right) &\mathrel{\widehat{=}}& z^{e,RPU}  \,,  \enspace \forall j \, , \label{eq:cost2}\\
                        \Rightarrow v_{OC,j}^{e,RPU}&\mathrel{\widehat{=}}&v_{OC,j}\left(t_{RPU}^{e}\right) &\mathrel{\widehat{=}}& v_{OC}^{e,RPU}  \,,  \enspace \forall j \, . \label{eq:cost3}     
                    \end{align} 
                \end{subequations} 
                
                Based on (\ref{eq:optim}), \(Q^{e,RPU}\) can now be calculated for both Approaches. For Approach 1, the intermittent goal is to reach two expressions for \(v_{OC}^{e,RPU}\): firstly, by starting from (\ref{eq:cost}); secondly, by starting from (\ref{eq:cost2}). The two expressions are subsequently equated. The resulting expression is solved numerically to produce \(Q^{e,RPU}\).
                
                Therefore, normalising (\ref{eq:cost}) by \(Q_{c,nom}\), applying (\ref{eq:PropFactor_b}) and denormalising by \(R_{c,nom}\) results in equivalence of cell resistances at \(t_{RPU}^{e}\), i.e.,
                \begin{equation}\label{eq:CellRes_EOL_PU_withrec}
                    R_{j}^{e,RPU} \mathrel{\widehat{=}} R_{j}\left(t_{RPU}^{e}\right) \mathrel{\widehat{=}} R^{e,RPU}  \,,  \quad  \forall j \, .
                \end{equation}  
                
                Furthermore, by applying (\ref{eq:cost3}) and (\ref{eq:CellRes_EOL_PU_withrec}) at \(t_{RPU}^{e}\), in line with Condition Remark~\ref{condrem:CondRem2} we have
                \begin{align}\label{eq:v_oc}
                    \sum_{j=1}^{N_p} v_{min} &= \sum_{j=1}^{N_p}\left( v_{OC}^{e,RPU} + i_{j}\left(t_{RPU}^{e}\right)R^{e,RPU} \right) \nonumber \\ 
                    N_p v_{min} &= N_p v_{OC}^{e,RPU} + \sum_{j=1}^{N_p}\left( i_{j}^{e,RPU}\right)R^{e,RPU}  \\
                    &\Rightarrow   i_{j}^{e,RPU} = \frac{i_{PU}^{1C}}{N_p}  \,,  \quad \forall j          \, . \nonumber
                \end{align}
                
                Rewriting (\ref{eq:v_oc}) with consideration of (\ref{eq:EFCfromQ_EOL}) and (\ref{eq:ResIncrease_tot}) gives the first expression for \(v_{OC}^{e,RPU}\):
                \begin{equation}\label{eq:OCV_eq_1}
                    v_{OC}^{e,RPU} = \alpha + \beta Q^{e,RPU} \, ,
                \end{equation}
                where
                \begin{subequations}
                    \begin{align}
                        \alpha &= v_{min} - i_{PU}^{1C}\left(c_{j} + a_{j}d_{j}\right)/ N_p \, , \label{eq:alpha} \\
                        \beta &= i_{PU}^{1C}d_{j} b_{j}/N_p  \, . \label{eq:beta} 
                    \end{align}
                \end{subequations}                
                
                Now, considering the last CC discharge cycle, starting at \(t_{dis,last}^{s}\) and finishing at \(t_{PU}^{e}\) (i.e., \(t_{dis,last}^{s} < t < t_{PU}^{e}\)), the standard SOC equation for a cell in the PU is 
                \begin{equation}\label{eq:Cell_SOC_eq}
                    z_j(t) = 1 + \frac{1}{Q_j(t)}\int_{t_{dis,last}^{s}}^{t}i_j(\tau)d\tau \,,  \quad \forall j   \,\, .    
                \end{equation}
                
                For the same cycle, 
                applying Coulomb counting over the RPU with consideration of (\ref{eq:PU_EOL_Method1}) gives
                \begin{equation}\label{eq:SummingCurrents}
                    \begin{aligned}
                        \sum_{j=1}^{N_p}\int_{t_{dis,last}^{s}}^{t_{RPU}^{e}}i_j(\tau)d\tau   &= \int_{t_{dis,last}^{s}}^{t_{RPU}^{e}}  \sum_{j=1}^{N_p}i_j(\tau)d\tau \\
                        &= \int_{t_{dis,last}^{s}}^{t_{RPU}^{e}}  i_{PU}^{1C}d\tau \\
                        &= -0.8 \,Q_{PU,nom}^{1C} \, .   
                    \end{aligned}                 
                \end{equation}

                Evaluating (\ref{eq:Cell_SOC_eq}) at \(t_{RPU}^{e}\), summing over \(N_p\), and applying (\ref{eq:SummingCurrents}) produces the second expression for \(v_{OC}^{e,RPU}\):
                \begin{equation}\label{eq:OCV_eq_2}
                    v_{OC}^{e,RPU} = f\left(z^{e,RPU}\right) = f\left(1 - \frac{0.8}{N_p} \frac{Q_{PU,nom}^{1C}}{Q^{e,RPU}}\right)  \, ,                 
                \end{equation}           
                where \(f\) is the OCV-SOC curve from Assumption~\ref{Ass:OCVSOCfixed}.   

                Finally, equating (\ref{eq:OCV_eq_1}) and (\ref{eq:OCV_eq_2}) results in
                \begin{equation}\label{eq:alphabeta}
                    \alpha + \beta  Q^{e,RPU} = f\left(1 - \frac{0.8}{N_p}\frac{Q_{PU,nom}^{1C}}{Q^{e,RPU}}\right) \,\, .
                \end{equation} 
        
                In (\ref{eq:alphabeta}), all parameters apart from \(Q_{PU,nom}^{1C}\) and \(Q^{e,RPU}\) are known. The former is obtained in line with (\ref{eq:Q_PU_nom}) and the latter by numerically solving (\ref{eq:alphabeta}).                 
                
                For Approach 2, \(Q^{e,RPU}\) is, in line with (\ref{eq:PU_EOL_Method2}), directly determined by setting \(Q^{e,RPU} = 0.8 \, Q_{c,nom}\). 
                
                For any Approach, the EFC of RPU at \(t_{RPU}^{e}\) is obtained by summing the cell EFCs at \(t_{RPU}^{e}\) and using~(\ref{eq:EFCfromQ_EOL}):
                \begin{equation}\label{eq:EFCfromQ_EOL_sum_withrec}
                    \begin{aligned}
                        EFC_{RPU}^{e} 
                        &= \sum_{j=1}^{N_p}EFC_{j}^{e,RPU} \\
                        &= \sum_{j=1}^{N_p}\left(a_{j}-b_{j}Q_{j}^{e,RPU}\right)                          
                    \end{aligned}
                \end{equation}                                

            \paragraph{PU lifetime extension}\label{par:PU_RecGain}
                Ultimately, linking (\ref{eq:EFC_EOL_noreconf}) and  (\ref{eq:EFCfromQ_EOL_sum_withrec}) allows the definition of the \emph{PU lifetime extension in experiment \(k\)}:                       
                \begin{equation}\label{eq:RecGain_k}
                    \chi_{PU}^{k}   = \left(\frac{EFC_{RPU}^{e}}{EFC_{FPU}^{e}} -1\right) \cdot 100 \,[\%]  \, ,
                \end{equation}
                where an experiment refers to the simulation and calculations corresponding to one realization of cell aging models (cf. Fig.~\ref{fig:ageingModels}). The \emph{statistics of PU lifetime extension} can be defined in terms of the sample mean \(\bar{\chi}_{PU}\) and the sample standard deviation \(s_{\chi_{PU}}\) of a set of \(N_{exp}^{PU}\) experiments, i.e. 
                \begin{subequations}\label{eq:RecGain}
                    \begin{align}
                        \bar{\chi}_{PU}   &= 
                        \frac{1}{N_{exp}^{PU}} 
                        \sum_{k=1}^{N_{exp}^{PU}} 
                        \chi_{PU}^{k} \label{eq:RecGain_mu}\\
                        s_{\chi_{PU}}   &= 
                        \sqrt{\frac{1}{N_{exp}^{PU}-1} 
                        \sum_{k=1}^{N_{exp}^{PU}}
                        \left(\chi_{PU}^{k} - \bar{\chi}_{PU}\right)^2} \label{eq:RecGain_sigma}           
                    \end{align}
                \end{subequations}

        \subsubsection{GM analysis}
            Since the same current passes through all PUs in the FGM, the FGM 
            reaches its EOL when the first of its series-connected FPUs 
            reaches its EOL. The \emph{GM lifetime extension in experiment \(k\)} is therefore
            \begin{equation}\label{eq:RecGain_GRM_singleCase}
                \chi_{GM}^{k}   
                = 
                \left(
                \frac{\frac{1}{N_s}\sum\limits_{i=1}^{N_s}EFC_{RPU,i}^{e}}
                {\min\limits_{i}\left({EFC_{FPU,i}^{e}}\right)} 
                -1
                \right) \cdot 100 \,[\%]                \, .
            \end{equation}
            Similarly to (\ref{eq:RecGain}), the \emph{statistics of GM lifetime extension} can be defined in terms of the sample mean \(\bar{\chi}_{GM}\) and the sample standard deviation \(s_{\chi_{GM}}\) of a set of \(N_{exp}^{GM}\) experiments:
            \begin{subequations}\label{eq:RecGain_GRM}
                \begin{align}
                    \bar{\chi}_{GM}   &= 
                    \frac{1}{N_{exp}^{GM}} 
                    \sum_{k=1}^{N_{exp}^{GM}} 
                    \chi_{GM}^{k} \label{eq:RecGain_GRM_mu}\\
                    s_{\chi_{GM}}   &= 
                    \sqrt{\frac{1}{N_{exp}^{GM}-1} 
                    \sum_{k=1}^{N_{exp}^{GM}}
                    \left(\chi_{GM}^{k} - \bar{\chi}_{GM}\right)^2} \label{eq:RecGain_GRM_sigma}                        
                \end{align}                
            \end{subequations}   

\begin{figure*}
  \centering
    \subfloat{\includegraphics[width=.32\textwidth]{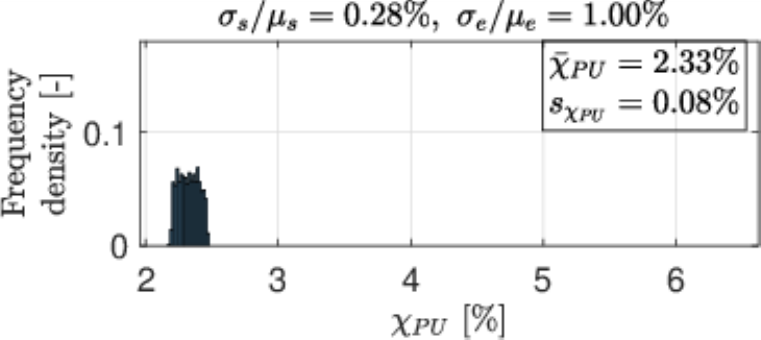}}\,\,
    \subfloat{\includegraphics[width=.32\textwidth]{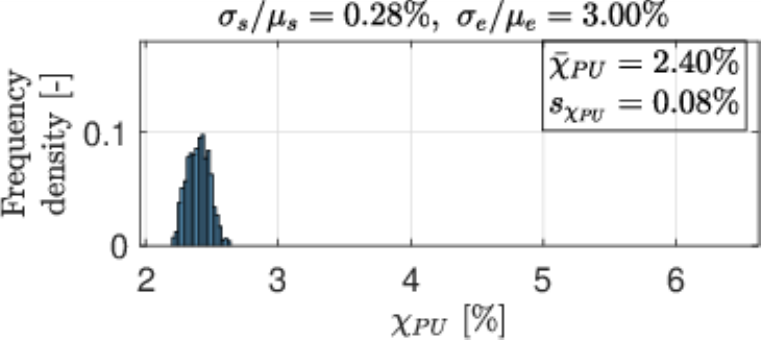}}\,\,
    \subfloat{\includegraphics[width=.32\textwidth]{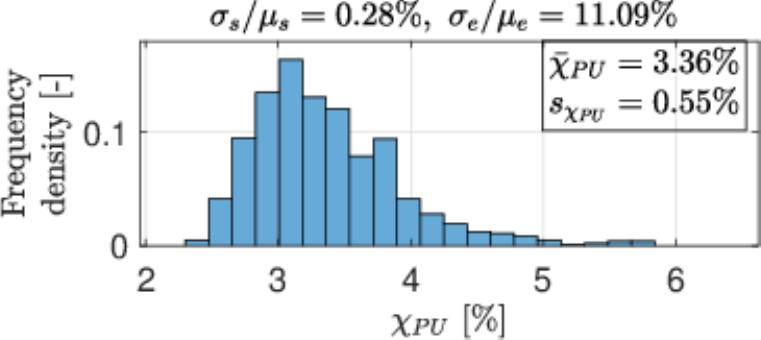}}\,\,
  \caption{Exemplified histograms of PU lifetime extension for the case of \(N_p = 10\) and \(\rho = 105.7 ^{\circ}\), evaluated using Approach 1, normalised over \(N_{exp}^{PU}\).
  }
  \label{fig:Tail}
\end{figure*}               

    \section{Implementation}\label{sec:Implementation}

    \subsection{Data}
    
        \subsubsection{Cell electrical model}
            The employed cell electrical model encompasses OCV, a series resistance \(R_0\) and a single RC pair (consisting of polarization resistance \(R_1\) and capacitance \(C_1\)). OCV is a function of SOC
            , while \(R_0\), \(R_1\), and \(C_1\) are 
            constant over the given temperature.  The cell model parameters originate from \cite{Ple:15-2}.

        \subsubsection{Cell ageing model}        
            The used fraction of the cell ageing dataset, originating from \cite{Li:21-1}, comprises data on 48 nickel-manganese-cobalt (NMC) cells. In \cite{Li:21-1}, the cells were cycled using 1C-discharge and 1C-charge current
            alongside periodic characterization measurements, yielding cell capacity fade data 
            as a function of EFC and cell resistance increase data 
            (characterized at \(50 \%\) SOC) 
            as a function of EFC. 

            Within this analysis, the cell capacity fade data from \cite{Li:21-1} are normalised and fit 
            to (\ref{eq:Q_distribution}) and (\ref{eq:EFC_distribution})  
            using least-squares, resulting in the mean \(\mu_{s,fit}\) and the standard deviation \(\sigma_{s,fit}\) for the fitted normal distribution at the BOL and the mean \(\mu_{e,fit}\) and the standard deviation \(\sigma_{e,fit}\) for the fitted normal distribution at the EOL:            \begin{equation}\label{eq:FittedBOLvalue}       
                \begin{split}
                    \widehat{\widetilde{Q}}_{c,s} &\sim \mathcal{N}\left(\mu_{s,fit},\,\sigma^{2}_{s,fit}\right) \, ,\\
                    \widehat{{EFC}}_{c,e} &\sim \mathcal{N}\left(\mu_{e,fit},\,\sigma^{2}_{e,fit}\right) \, ,   
                \end{split}
            \end{equation} 
            with 
            \begin{equation}
                \begin{split}
                \{\mu_{s,fit}, \sigma_{s,fit}\}   &= \{0.9939, 0.0028 \} \, ,\\ 
                \{\mu_{e,fit}, \sigma_{e,fit} \} &= \{ 615.85, 68.28\}  \, .               
                \end{split}             
            \end{equation}

            Moreover, within this analysis, the cell resistance increase data from \cite{Li:21-1} are not directly used. Instead, the best least-squares fit is used to obtain \(k_{RQ}\) and \(l_{RQ}\) (cf. (\ref{eq:PropFactor_b}) and Fig.~\ref{fig:ageingModel_dQ_dR}).

    \subsection{Experiment framework}

        Within this analysis, \texttt{SimScape} is used for cell modeling, while \texttt{MATLAB} and \texttt{Simulink} are used for conducting simulation experiments on the Linux-based computer cluster with 24 cores.
        
        \subsubsection{PU experiments}\label{subsubsec:PUcase}
        
            The parameters deemed to be able to influence the PU lifetime extension significantly are linked to cell manufacturing variations (\(\sigma_{s}\)), cell chemistry and usage characteristics (\(\sigma_{e}\)), cell chemical properties (\(\rho\)), and PU configuration (\(N_p\)).
            
            Concerning \(\sigma_{s}\) and \(\sigma_{e}\), two additional representative perturbations are selected alongside \(\sigma_{s,fit}\) and \(\sigma_{e,fit}\): 
            \begin{subequations}\label{eq:BOL_EOL_array}
                \begin{align}
                    \frac{\sigma_{s}}{\mu_{s,fit}} &\in \{0.1\%, 0.28\% \,\text{(fit.)}, 1\%\} \, ,\label{eq:BOL_array} \\
                    \frac{\sigma_{e}}{\mu_{e,fit}}  &\in \{1\%, 3\%, 11.1\% \,\text{(fit.)}\} \, .\label{eq:EOL_array}               
                \end{align}
            \end{subequations}

            Moreover, two additional representative perturbations of \(\rho\) are selected alongside \(\rho_1 \mathrel{\widehat{=}} \rho_{fit}\), evaluated using (\ref{eq:rho}), and depicted in Fig.~\ref{fig:ageingModel_dQ_dR}:
            \begin{equation}\label{eq:rho_array}
                \rho \in \{124.5 ^{\circ} \text{(fit.)} , 105.7 ^{\circ}, 97.3 ^{\circ}\} \, .      
            \end{equation}            
    
            Finally, regarding \(N_p\), the following values are selected: 
            \begin{equation}\label{eq:N_p_array}
                N_p \in \{2, 4, 6, 8, 10, 12, 20\} \, .
            \end{equation}         
            
            Hence, the dimension of the simulation setup was \(3^3 \times 7 = 189\) cases. The Monte Carlo method with \(N_{exp}^{PU} = 1000\) experiments is applied for every case. At the start of every experiment, the cells have the same initial SOC of \(50 \%\).   
            For every experiment, FPU performance is evaluated by using (\ref{eq:PU_EOL_Method1})--(\ref{eq:EFC_EOL_noreconf}), while RPU performance is obtained by means of (\ref{eq:optim})--(\ref{eq:EFCfromQ_EOL_sum_withrec}). Finally, statistics of PU lifetime extension is calculated on the basis of (\ref{eq:RecGain_k}) and (\ref{eq:RecGain}).

        \subsubsection{GM experiments}    
            An additional factor deemed to be able to influence the GM lifetime extension significantly is linked to GM configuration (\(N_s\)).                        
            Regarding \(N_s\), the following values are selected: 
            \begin{equation}\label{eq:N_s_array}
                N_s \in \{2, 3, \dots, 9, 10, 15, \dots, 200\} \, ,
            \end{equation}                  
            with the last entry reflecting the \(800 \, \text{V}\) battery system.

            Within a single case, for each \(N_s\), a random sample of indices of size \(N_s\) is selected, with indices ranging between 1 and \(N_{exp}^{PU}\). Performance indicators (\ref{eq:EFC_EOL_noreconf}) and (\ref{eq:EFCfromQ_EOL_sum_withrec}) corresponding to the selected indices are obtained. This procedure is repeated \(N_{exp}^{GM} = 100000\) times. Statistics of GM lifetime extension is subsequently obtained using (\ref{eq:RecGain_GRM_singleCase}) and (\ref{eq:RecGain_GRM}).      

    \section{Results And Discussion}\label{sec:Results} 

    Results for the PU lifetime extension \(\chi_{PU}\)
    are, for all the 189 cases, obtained in terms of histograms of \(\chi_{PU}\), as exemplified in Fig.~\ref{fig:Tail}. In order to facilitate the comparison between different cases, the corresponding \(\bar{\chi}_{PU}\) and \(s_{\chi_{PU}}\) are also calculated for every histogram.    
    
\begin{figure*}\sffamily       
        \begin{tabular}{l@{\hspace{0.5pt}}C@{\hspace{0.5pt}}C@{\hspace{0.5pt}}C@{\hspace{0.5pt}}C@{\hspace{0.5pt}}}
            \toprule
                Nr. & a & b & c & d \\ 
            \midrule
                \rotatebox[origin=c]{90}{A1m} & \addpic{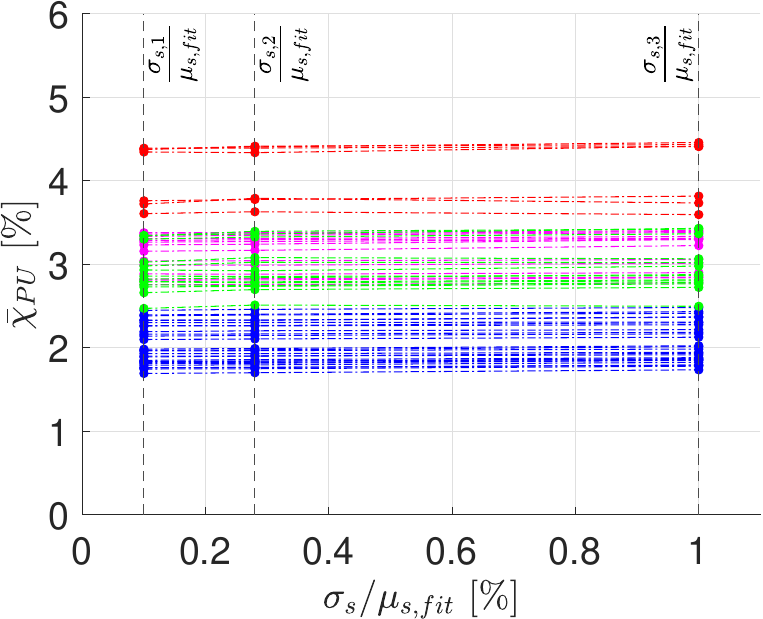} & \addpic{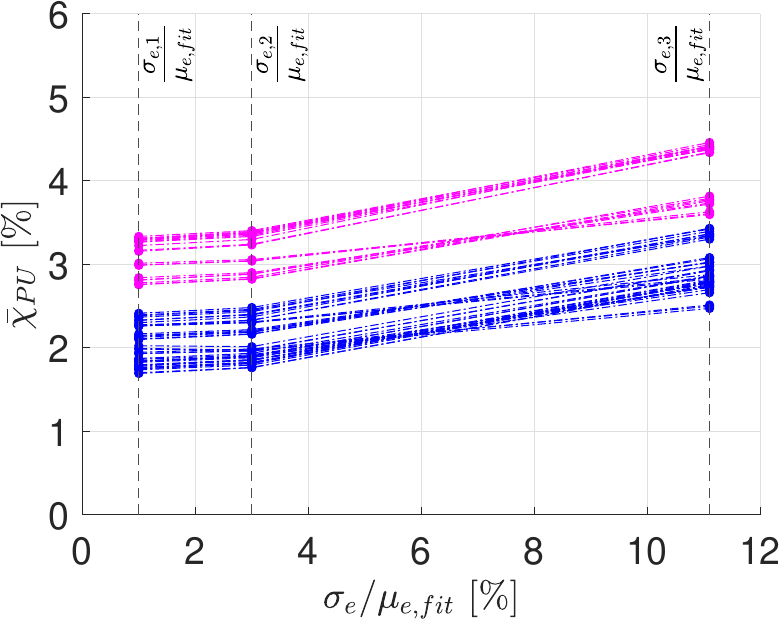} & \addpic{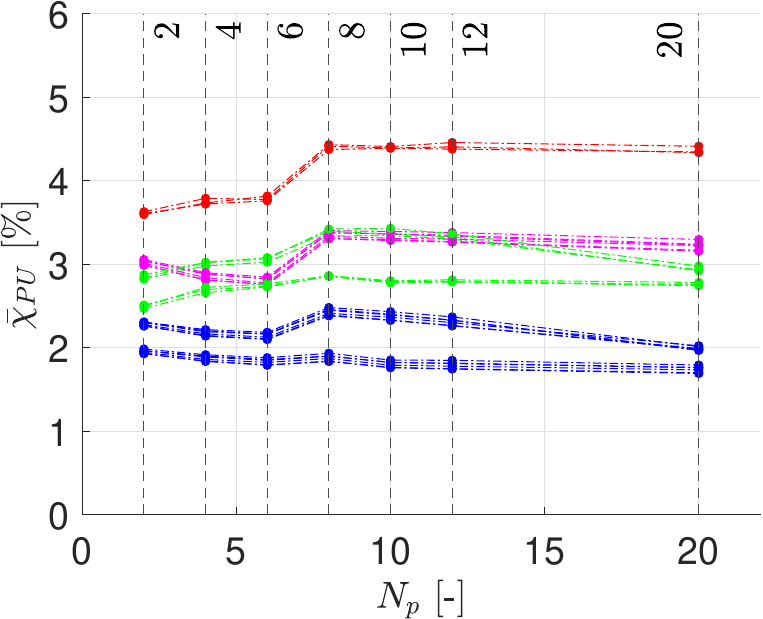} & \addpic{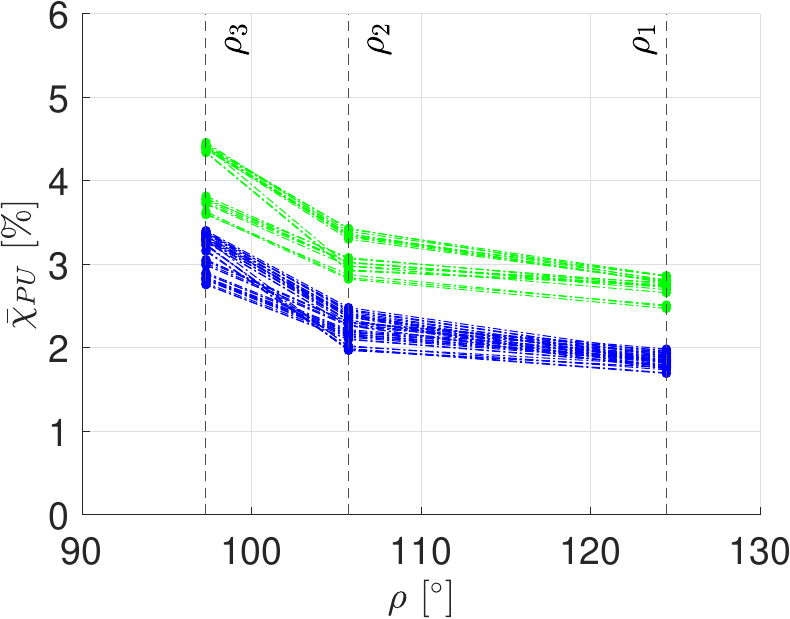} \\ 
                \rotatebox[origin=c]{90}{A1s} & \addpic{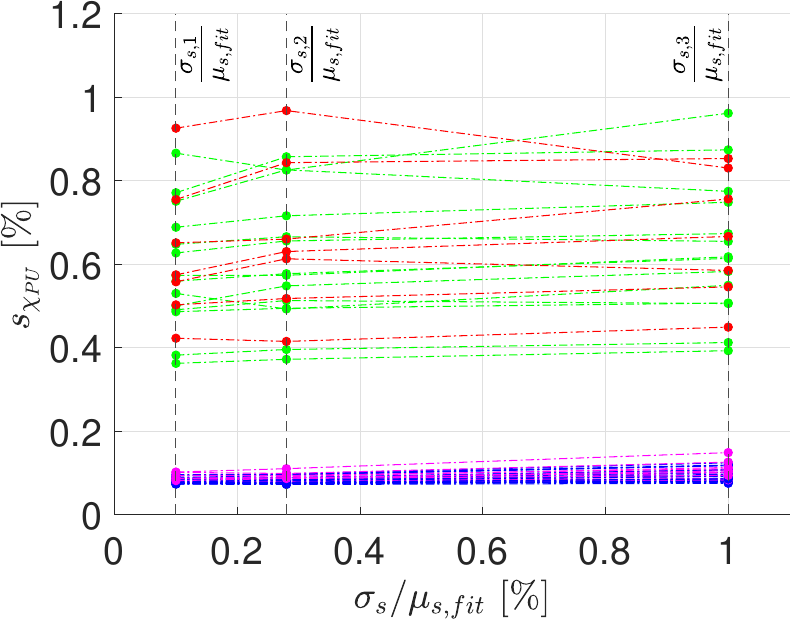} & \addpic{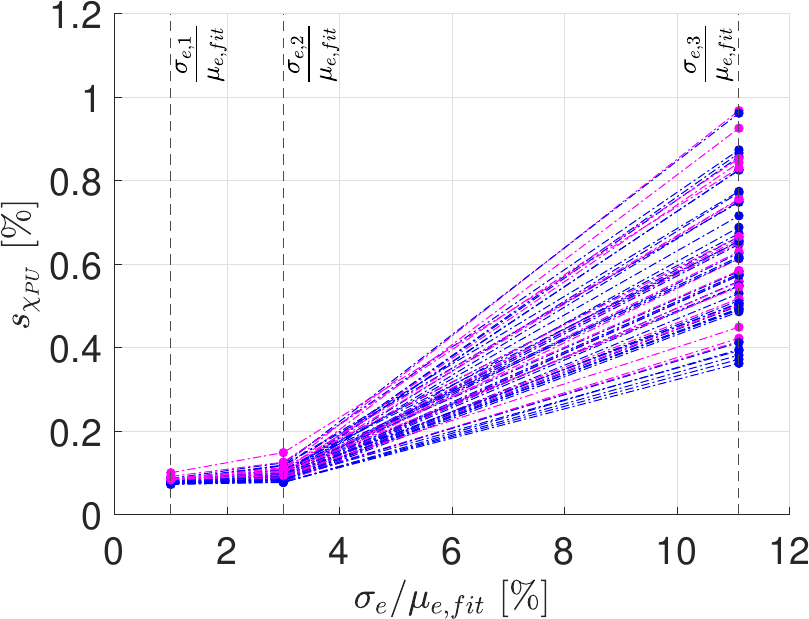} & \addpic{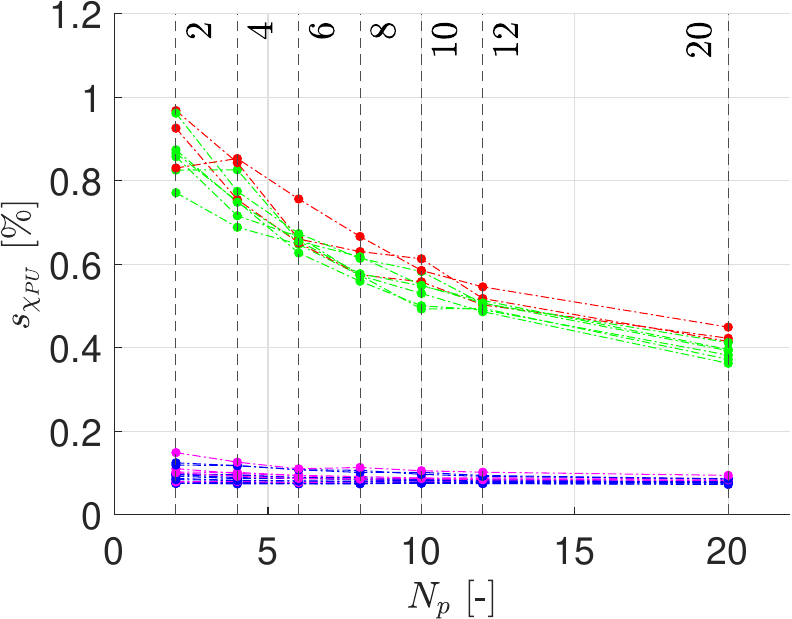} & \addpic{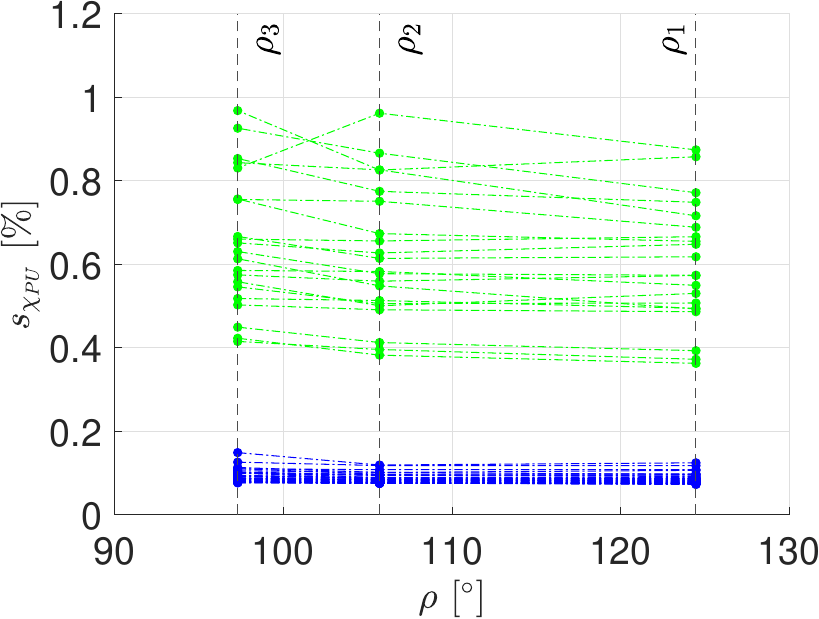} \\ 
                \rotatebox[origin=c]{90}{A2m} & \addpic{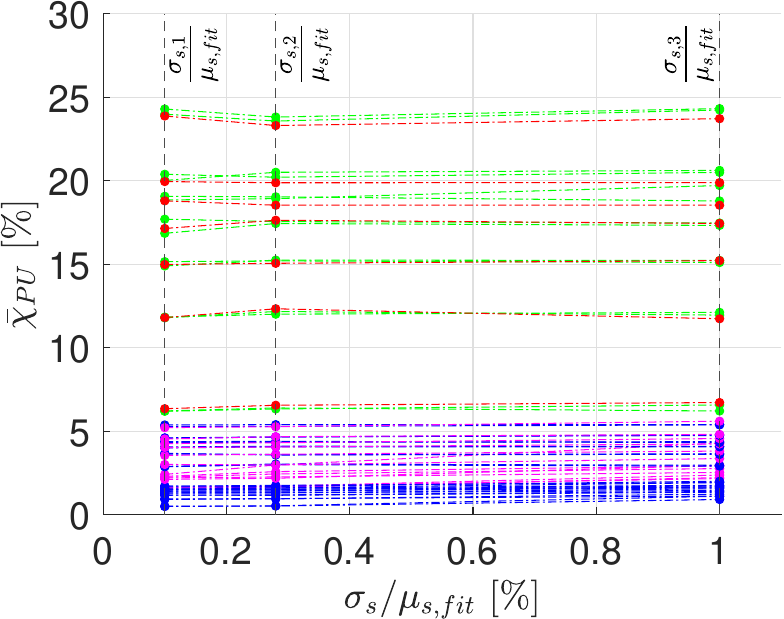} & \addpic{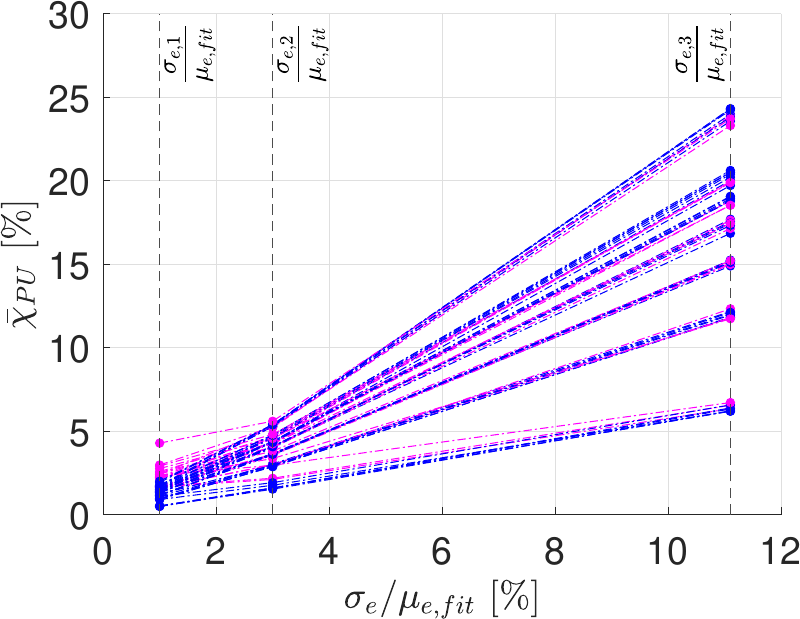} & \addpic{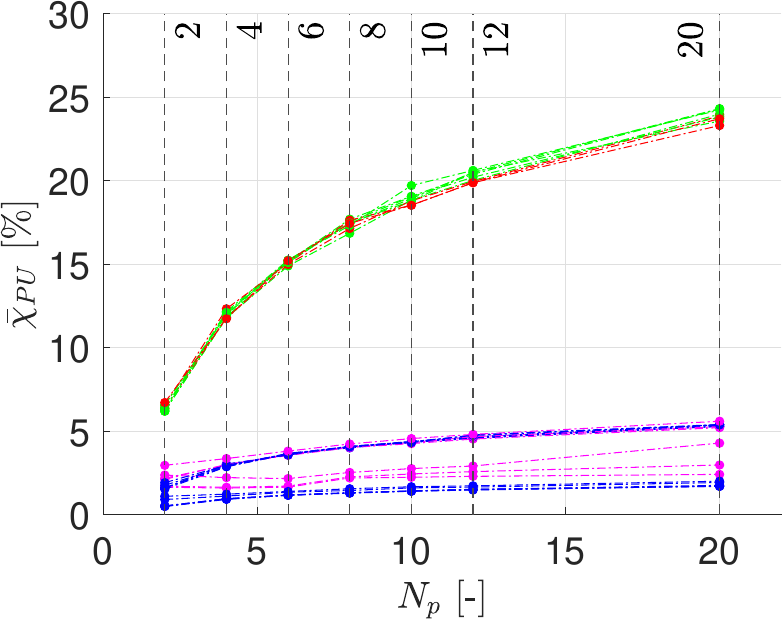} & \addpic{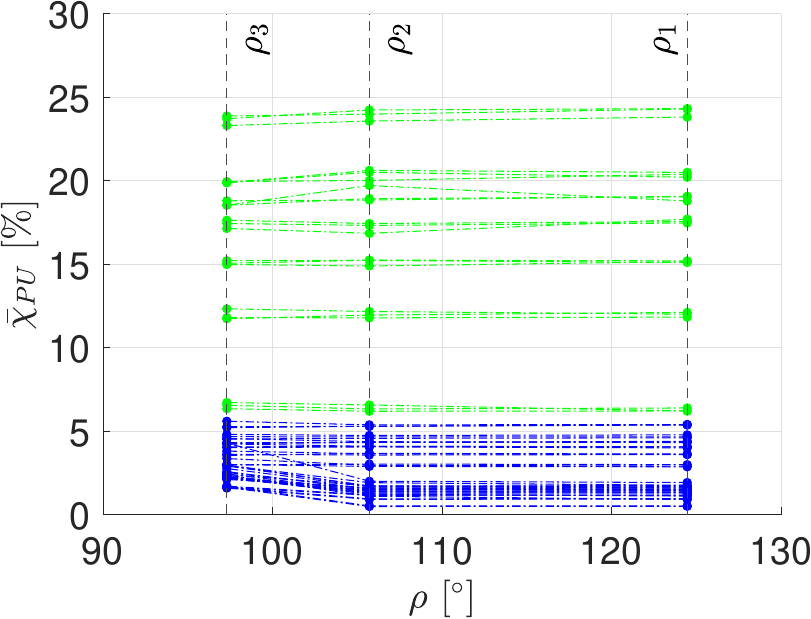} \\ 
                \rotatebox[origin=c]{90}{A2s} & \addpic{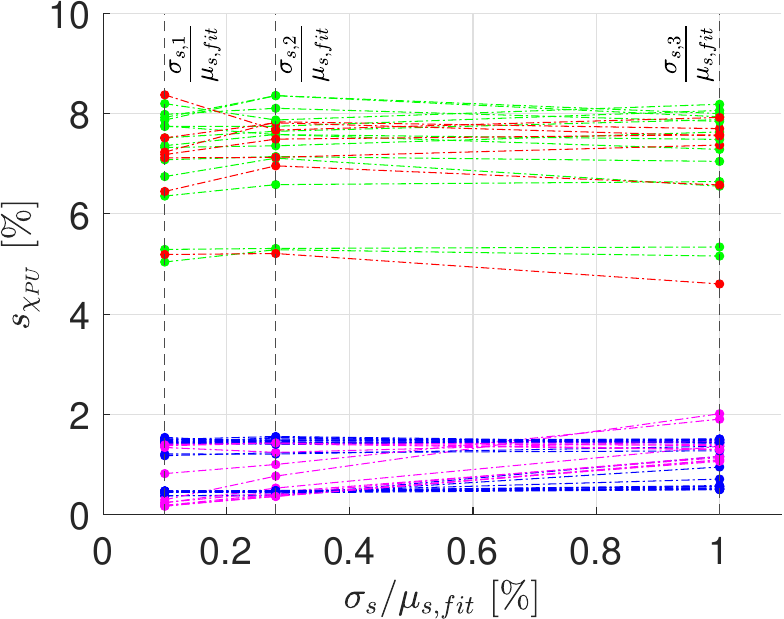} & \addpic{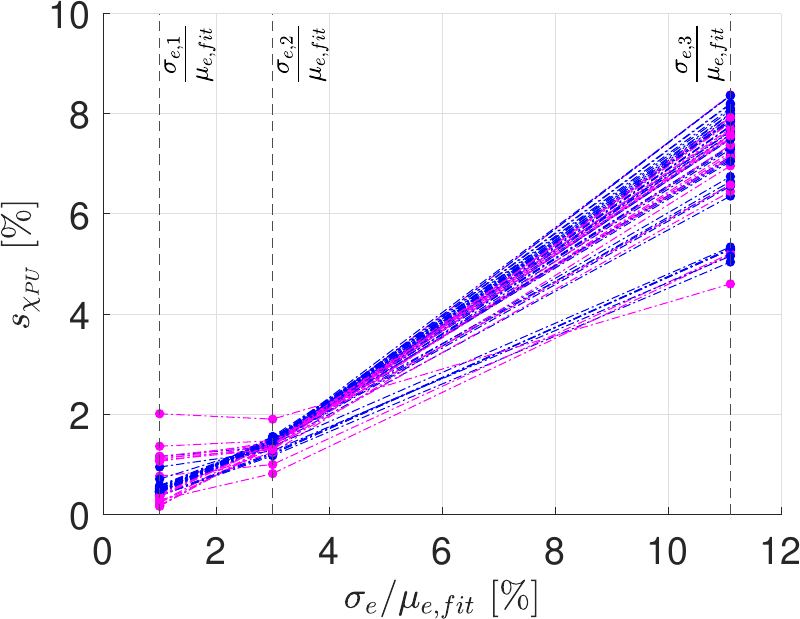} & \addpic{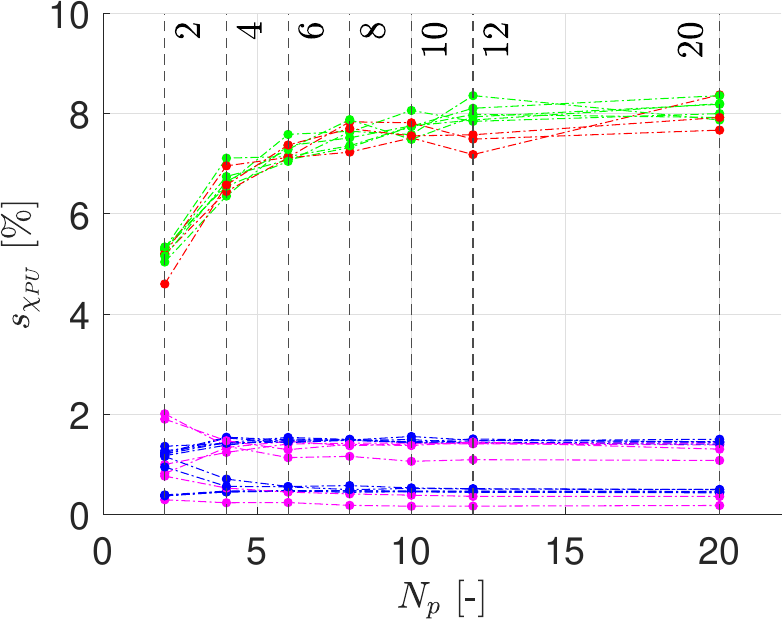} & \addpic{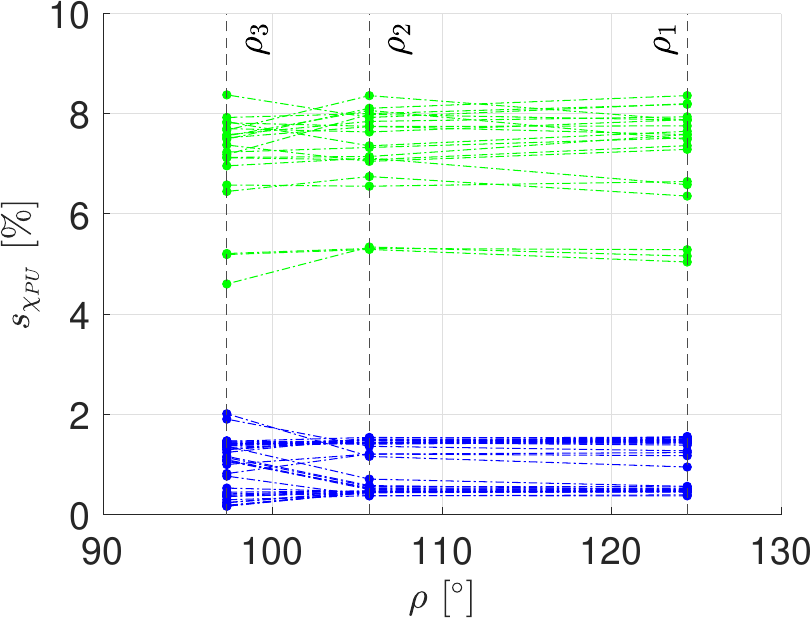} \\ 
            \bottomrule 
        \end{tabular}
        \caption{Statistics of PU lifetime extension. Influence of perturbation of a) cell capacity spread at the BOL, b) cell EFC spread at the EOL, c) number of cells in the PU, d) cell resistance increase to cell capacity fade relationship. Influence on the mean (A1m and A2m) and the standard deviation (A1s and A2s) of the PU lifetime extension, evaluated for Approach 1 (A1m and A1s) and 2 (A2m and A2s). Green: data for \(\sigma_{e} = \sigma_{e,3}\). Magenta: data for \(\rho = \rho_3\). Red: data for both \(\sigma_{e} =\sigma_{e,3}\) and \(\rho =\rho_3\). Blue: remaining data.}     
        \label{fig:BigFigure}
    \end{figure*}

\begin{figure*}\sffamily  
            \begin{tabular}{l@{\hspace{0.5pt}}E@{\hspace{0.5pt}}E@{\hspace{0.5pt}}}
                \toprule
                    Nr. & a & b \\ 
                \midrule
                    \rotatebox[origin=c]{90}{A1} & 
                    \addpec{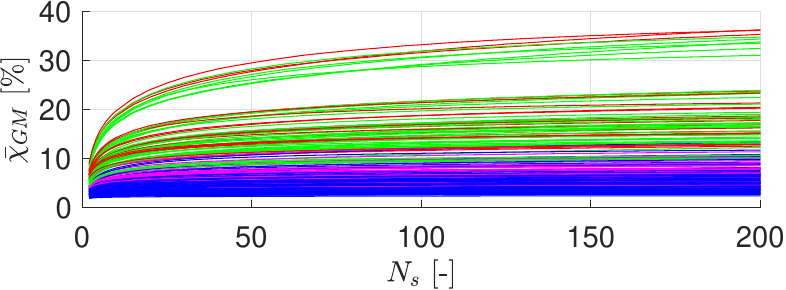} & \addpec{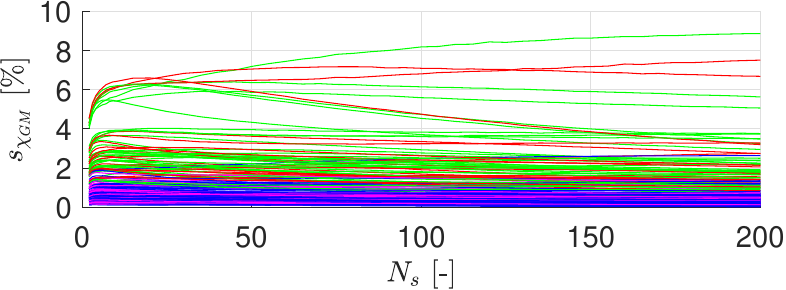} \\ 
                    \rotatebox[origin=c]{90}{A2} & \addpec{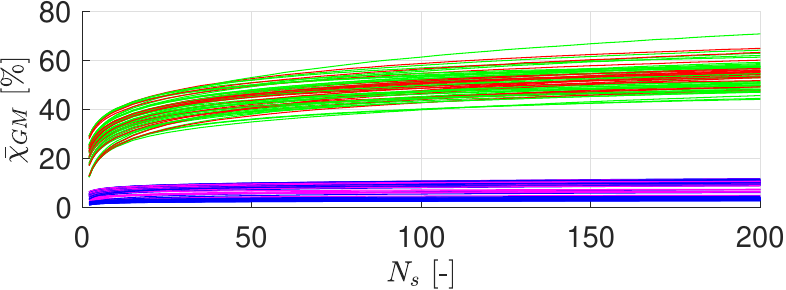} & \addpec{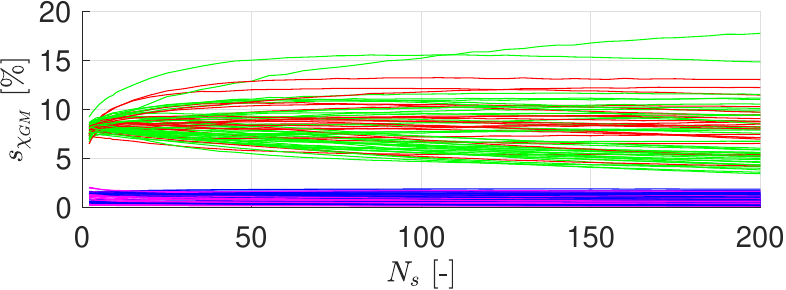} \\ 
                \bottomrule 
            \end{tabular}        
            \caption{Statistics of GM lifetime extension. Influence of the number of series-connected PUs on the mean (a) and the standard deviation (b) of the GM lifetime extension, evaluated for Approach 1 (A1) and 2 (A2). Green: data for \(\sigma_{e} = \sigma_{e,3}\). Magenta: data for \(\rho = \rho_3\). Red: data for both \(\sigma_{e} =\sigma_{e,3}\) and \(\rho =\rho_3\). Blue: remaining data.}     
            \label{fig:BigFigure2}
        \end{figure*}

    The influence of perturbation of the individual parameters (\ref{eq:BOL_EOL_array})--(\ref{eq:N_p_array}) on \(\chi_{PU}\) is presented in Fig.~\ref{fig:BigFigure}.
    Every data point in Fig.~\ref{fig:BigFigure} 
    corresponds to one of the 189 cases.
    Analogously, the results for \(\chi_{GM}\) are, for the 189 cases and the 47 \(N_s\) values from (\ref{eq:N_s_array}), obtained in terms of histograms of \(\chi_{GM}\) and the corresponding statistics \(\bar{\chi}_{GM}\) and \(s_{\chi_{GM}}\). The influence of perturbation of \(N_s\) on \(\chi_{GM}\) is illustrated in Fig.~\ref{fig:BigFigure2}. In contrast to 
    Fig.~\ref{fig:BigFigure}, 
    every line in Fig.~\ref{fig:BigFigure2} corresponds to one of the 189 cases. 

    \subsection{Lifetime extension of RBSs}

        RBS has a significant potential to extend the lifetime of the battery pack systems with a fixed cell configuration (FBS) which are commonly used today. This is corroborated by each of the 189 investigated cases, as partially illustrated in Fig.~\ref{fig:Tail} where \(\chi_{PU}\) is positive in all experiments. Furthermore, the lifetime extension is particularly pronounced for the battery packs consisting of cells with a large EFC spread at the EOL and for battery packs comprising a large number of series-connected elements (namely large values of \(\sigma_{e}\) and \(N_s\)). 
 
        When evaluated using the safety-based Approach 2, \(\bar{\chi}_{PU}\) is between \(0.48 \, \%\) and \(24.31 \, \%\), depending on the investigated case. Series-connecting the corresponding PUs results in an even higher lifetime extension. Specifically, for \(N_s = 200\), corresponding to the \(800 \, \text{V}\) battery system, the expected lifetime extension \(\bar{\chi}_{GM}\) ranges between \(2.62 \, \%\) and \(70.84 \, \%\).    

        When evaluated using capacity-based Approach 1, \(\bar{\chi}_{PU}\) ranges between \(1.69\,\%\) and \(4.46\,\%\). Correspondingly, for the \(800 \, \text{V}\) battery system with \(N_s = 200\), the highest expected lifetime extension is as large as \(36.25 \, \%\). Although inferior to the result obtained by evaluation for Approach 2, it is still a significant improvement over FBS. The difference in lifetime extension between the two Approaches is a strong indicator of the relevance of the dynamic battery reconfiguration not only for the lifetime extension of battery packs, but for their safety as well. 

    \subsection{Effects of PU parameters}
      
        \subsubsection{\texorpdfstring{Effects of \(\sigma_{s}\)}{Effects of sigma s}}
            Perturbations of \(\sigma_{s}\), evaluated using both Approaches, show a marginal effect on \(\chi_{PU}\) as both \(\bar{\chi}_{PU}\) and \(s_{\chi_{PU}}\) change insignificantly for an increase in \(\sigma_s\)  (cf. Fig.~\ref{fig:BigFigure}a).

        \subsubsection{\texorpdfstring{Effects of \(\sigma_{e}\)}{Effects of sigma e}}
            In contrast to the marginal effect of \(\sigma_{s}\), perturbations of \(\sigma_{e}\), evaluated using both Approaches, have a pronounced effect on \(\chi_{PU}\).
            An increase in \(\sigma_e\) results in considerable increases in both \(\bar{\chi}_{PU}\) and \(s_{\chi_{PU}}\) (cf. Fig.~\ref{fig:BigFigure}b), in particular for Approach 2 where the increase is almost tenfold.  
        
            The PU lifetime extension in a single experiment \(\chi_{PU}^{k}\) is illustratively determined by the relationship between black and red crosses in Fig.~\ref{fig:EOL_PU}. In light of Fig.~\ref{fig:ageingModel_dQ} and Fig.~\ref{fig:ageingModels}, it is reasonable to expect that perturbations of \(\sigma_s\) may affect the cell capacity fade trajectories less than perturbations of \(\sigma_e\). Moreover, for increased \(\sigma_e\), the cell EFC spread at the EOL of PU is wider, yielding a higher probability of one (or more) cell(s) in the PU having a cell capacity fade trajectory significantly different from the remaining cells. On the one hand, the cell capacity fade trajectories with significantly lower cell EFC at the EOL of PU than the remaining cells increase the risk of FPU reaching its EOL prematurely, thus rendering a higher \(\bar{\chi}_{PU}\), especially for Approach 2. On the other hand, the cell capacity fade trajectories with significantly higher cell EFC at the EOL of PU than the remaining cells improve the FBS lifetime and reduce the effect of  \(\chi_{PU}\), yielding a wider histogram of \(\chi_{PU}^{k}\) (i.e., higher \(s_{\chi_{PU}}\)). 

            It is worth noting that, as \(\sigma_e\) increases, the histograms of \(\chi_{PU}\) exhibit a right skew-symmetric behaviour with a pronounced \enquote{tail}, as exemplified in Fig.~\ref{fig:Tail}. This is particularly important for \(\chi_{GM}\). The more pronounced the \enquote{tail} is, the higher the probability that the nominator in (\ref{eq:RecGain_GRM_singleCase}) is higher, yielding a more pronounced GM lifetime extension.     
            The considerable effect of \(\sigma_e\) on \(\chi_{PU}\) mirrors the importance of the usage pattern of the RPU as a means of counteracting the variations in the cell manufacturing process and local temperature variations. In other words, using a proper control strategy, dynamic battery reconfiguration can be used to ameliorate the initial discrepancies between the cells and to serve as an effective long-term capacity balancing strategy.

        \subsubsection{\texorpdfstring{Effects of \(\rho\)}{Effects of rho}}
            Perturbations of \(\rho\), evaluated using the two Approaches, result in different tendencies concerning \(\chi_{PU}\). For Approach 1, an increase in \(\rho\) results in a clear decrease of \(\bar{\chi}_{PU}\) (cf. Fig.~\ref{fig:BigFigure}d), while the effect on \(s_{\chi_{PU}}\) is negligible. Hence, an increase in \(\rho\) shifts the histogram of \(\chi_{PU}\) to the right. For Approach 2, though, the perturbations in \(\rho\) have a negligible effect on both \(\bar{\chi}_{PU}\) and \(s_{\chi_{PU}}\). 
        
            The influence of the parameter \(\rho\) on \(\chi_{PU}\) reflects the resistance variations between the individual cells. Although perhaps unexpected at first, this influence can be linked to the PU 1C-capacity, which is used for Approach 1 (cf. (\ref{eq:PU_EOL_Method1})). As lower values of \(\rho\) correspond to higher cell resistance values for a given cell capacity (cf. Fig.~\ref{fig:ageingModel_dQ_dR}), the PU 1C-capacity is reduced accordingly due to the voltage drop over the resistance. The counter-part evaluation of \(\chi_{PU}\) for Approach~2 uses the true cell capacity (cf. (\ref{eq:PU_EOL_Method2})) and is not affected by this.

        \subsubsection{\texorpdfstring{Effects of \(N_p\)}{Effects of N p}}
            Finally, perturbations of \(N_p\), evaluated using the two Approaches, also result in different tendencies concerning \(\chi_{PU}\). For Approach 1, while the increase in \(N_p\) reduces or significantly reduces \(s_{\chi_{PU}}\), no uniform trend for \(\bar{\chi}_{PU}\) can be observed. An increase in \(N_p\) basically shrinks the histogram of \(\chi_{PU}\). In contrast to this, for Approach 2, an increase in \(N_p\) results in a clear increase of \(\bar{\chi}_{PU}\) and a marginal effect on \(s_{\chi_{PU}}\), apart from the cases which involve the highest \(\sigma_e\)-value from (\ref{eq:EOL_array}), for which \(s_{\chi_{PU}}\) increases as well.
        
            An increase in the number of cells in the PU while keeping the same \(\sigma_s\), \(\sigma_e\), and \(\rho\) corresponds to an increase in the number of cell capacity fade trajectories in Fig.~\ref{fig:ageingModels} while keeping the same \(\widetilde{Q}_{c,s}\) and \(EFC_{c,e}\), which can be interpreted as more dense \enquote{sampling} within a single case. In light of Approach 1, an increase in \(N_p\) may result in improved averaging of the EFC of FPU, yielding only small changes in \(\bar{\chi}_{PU}\) but a more accurate distribution (i.e., reduced \(s_{\chi_{PU}}\)). Possibly, the histograms of \(\chi_{PU}\) would eventually converge by further increasing \(N_p\). In contrast, for Approach 2, evaluation shows that an increase in \(N_p\) considerably affects \(\chi_{PU}\), which can be explained in a manner analogous to that of an increase in \(\sigma_e\).
        
            The influence of perturbations of parameter \(N_p\) on \(\chi_{PU}\) reflects the trade-off between the algorithm design and the PU configuration. On the one hand, provided that the commonly used 1C-capacity measurement is used for BMS algorithms, increasing the number of cells in the PU is not followed by the increased lifetime extension through dynamic battery reconfiguration. Increasing the number of cells in the PU may, however, lead to safety hazard 
            (cf. Fig.~\ref{fig:EOL_PU} where one aged cell could operate below the safety threshold). On the other hand, provided that the cell capacities are perfectly known, the dynamic battery reconfiguration can maximize the capacity potential of the battery, which is especially pronounced when a cell whose capacity fade trajectory differs significantly from the other cells is added to the PU.

    \subsection{Effects of GM parameters}\label{subsec:GRMlifext}       

        For all values of \(N_s\) from (\ref{eq:N_s_array}), both Approaches show that an increase in \(N_s\) results in an increase of \(\bar{\chi}_{GM}\) (cf. Fig.~\ref{fig:BigFigure2}a), which reflects the higher probability that some of the series-connected PUs will include cells which reach their EOL prematurely.             
        No uniform trend for  \(s_{\chi_{GM}}\) could be observed though (cf. Fig.~\ref{fig:BigFigure2}b). 
        The reason for this stems from the underlying statistics of (\ref{eq:RecGain_GRM_singleCase}). The distribution of the resulting statistics (\ref{eq:RecGain_GRM}) depends on the specific characteristics of the underlying finite-sample distribution of 
        (\ref{eq:RecGain_GRM_singleCase}).    

    \section{Conclusion}\label{sec:Conclusion}

    The field of RBSs is promising yet deficiently explored. Various researchers have tried to provide solutions that improve the RBS operation over FBS in a certain regard. This paper presents the first simulation-based analysis of the potential lifetime extension through dynamic battery reconfiguration. The system under study comprises series-connected units of parallel cells, where each such unit can be bypassed and each cell can be disconnected. A methodology to estimate the potential extension has been provided based on experimentally motivated assumptions on cell capacity fade, stochasticity, cell resistance increase, expressions derived for ideal reconfiguration, and extensive simulations.

    The lifetime extension with respect to several parameters has been analyzed and discussed for a publicly available dataset from cell ageing experiments on NMC Li-ion cells. It has been shown that, for the investigated cell type, the lifetime extension for the \(800 \, \text{V}\) battery system can be as high as \(71 \%\). The results indicate that the cell EFC distribution at the EOL and the number of series-connected units significantly impact the lifetime extension. Furthermore, the number of cells in parallel and the resistance increase with age can be influential as well, depending on the employed metrics.       
    
    \printbibliography

@article{zhang2022machine,
    title={A machine learning-based framework for online prediction of battery ageing trajectory and lifetime using histogram data},
    author={Zhang, Yizhou and Wik, Torsten and Bergstr{\"o}m, John and Pecht, Michael and Zou, Changfu},
    journal={J. Power Sources},
    volume={526},
    pages={231110},
    year={2022},
    publisher={Elsevier}
}

@article{Kane:12,
    author  = {Kane, Mark},
    date    = {2021-10-13},
    title   = {Watch {Tesla} {Model} {S} {Plaid}'s {Battery} {Get} {Opened} {And} {Described}},
    journal = {InsideEVs: Electric Vehicle News, Reviews, and Reports},
    url     = {https://insideevs.com/news/540380/tesla-models-plaid-battery-open/},
    urldate = {2023-02-01}
}

@software{Ple:15-2,
    author  = {Plett, Gregory},
    title   = {{ESC} Model Toolbox},
    url     = {http://mocha-java.uccs.edu/BMS1/CH02/ESCtoolbox.zip},
    urldate = {2023-02-15}
}

@article{Bouchhima:17,
    title = {Optimal energy management strategy for self-reconfigurable batteries},
    journal = {Energy},
    volume = {122},
    pages = {560-569},
    year = {2017},
    author = {Nejmeddine Bouchhima and Marc Schnierle and Sascha Schulte and Kai Peter Birke}
}

@article{Muhammad:19,
    author = {Muhammad, Shaheer and Rafique, M. Usman and Li, Shuai and Shao, Zili and Wang, Qixin and Liu, Xue},
    title = {Reconfigurable Battery Systems: A Survey on Hardware Architecture and Research Challenges},
    year = {2019},
    publisher = {Association for Computing Machinery},
    address = {New York, NY, USA},
    volume = {24},
    number = {2},
    journal = {ACM Trans. Des. Autom. Electron. Syst.},
    articleno = {19},
    numpages = {27}
}

@article{Sve:20,
    author  = {{Volkswagen group}},
    date    = {2020-10-02},
    title   = {Scania researches better batteries},
    url     = {https://www.volkswagenag.com/en/news/2020/10/scania_research_batteries.html},
    urldate = {2023-02-10}
}

@ARTICLE{9300283,
    author={Han, Weiji and Wik, Torsten and Kersten, Anton and Dong, Guangzhong and Zou, Changfu},
    journal={IEEE Industrial Electronics Magazine}, 
    title={Next-Generation Battery Management Systems: Dynamic Reconfiguration}, 
    year={2020},
    volume={14},
    number={4},
    pages={20-31}
}

@INPROCEEDINGS{Pinter:21,
    author={Pinter, Zoltan Mark and Papageorgiou, Dimitrios and Rohde, Gunnar and Marinelli, Mattia and Træholt, Chresten},
    booktitle={2021 56th International Universities Power Engineering Conference (UPEC)}, 
    title={Review of Control Algorithms for Reconfigurable Battery Systems with an Industrial Example}, 
    year={2021}
}

@INPROCEEDINGS{Huang:21,
    author={Huang, Xiaoliang and Jiang, Bowen and Liu, Yujing},
    booktitle={2021 IEEE 19th International Power Electronics and Motion Control Conference (PEMC)}, 
    title={A Reconfigurable Battery Supercapacitor Hybrid Energy System with Active Balancing for Vehicle Applications}, 
    year={2021}
}

@software{Li:21-1,
    author  = {Li, Weihan and Sengupta, Neil and Dechent, Philipp and Howey, David and Annaswamy, Anuradha and Sauer, Dirk Uwe},
    title   = {Battery degradation prediction datasets and codes at {ISEA}},
    url     = {https://git.rwth-aachen.de/isea/battery-degradation-trajectory-prediction},
    urldate = {2023-02-01}
}

@article{PAUL2013642,
    title = {Analysis of ageing inhomogeneities in lithium-ion battery systems},
    journal = {Journal of Power Sources},
    volume = {239},
    pages = {642-650},
    year = {2013},
    author = {Sebastian Paul and Christian Diegelmann and Herbert Kabza and Werner Tillmetz}
}

@article{Preger_2020,
    year = {2020},
    volume = {167},
    number = {12},
    author = {Yuliya Preger and Heather M. Barkholtz and Armando Fresquez and Daniel L. Campbell and Benjamin W. Juba and Jessica Romàn-Kustas and Summer R. Ferreira and Babu Chalamala},
    title = {Degradation of Commercial Lithium-Ion Cells as a Function of Chemistry and Cycling Conditions},
    journal = {Journal of The Electrochemical Society}
}

@article{PASTORFERNANDEZ2016574,
    title = {A Study of Cell-to-Cell Interactions and Degradation in Parallel Strings: Implications for the Battery Management System},
    journal = {Journal of Power Sources},
    volume = {329},
    pages = {574-585},
    year = {2016},
    author = {C. Pastor-Fernández and T. Bruen and W.D. Widanage and M.A. Gama-Valdez and J. Marco}
}

@INPROCEEDINGS{Engelhardt,
    author={Engelhardt, Jan and Zepter, Jan Martin and Gabderakhmanova, Tatiana and Marinelli, Mattia},
    booktitle={2022 International Power Electronics Conference}, 
    title={Efficiency Characteristic of a High-Power Reconfigurable Battery with Series-Connected Topology}, 
    year={2022},
    volume={},
    number={},
    pages={2370-2376}
}

@Article{DoubleString,
    AUTHOR = {Engelhardt, Jan and Zepter, Jan Martin and Gabderakhmanova, Tatiana and Rohde, Gunnar and Marinelli, Mattia},
    TITLE = {Double-String Battery System with Reconfigurable Cell Topology Operated as a Fast Charging Station for Electric Vehicles},
    JOURNAL = {Energies},
    VOLUME = {14},
    YEAR = {2021},
    NUMBER = {9},
    ARTICLE-NUMBER = {2414}
}

@article{SONG2021100091,
    title = {A study of cell-to-cell variation of capacity in parallel-connected lithium-ion battery cells},
    journal = {eTransportation},
    volume = {7},
    pages = {100091},
    year = {2021},
    author = {Ziyou Song and Xiao-Guang Yang and Niankai Yang and Fanny Pinto Delgado and Heath Hofmann and Jing Sun}
}

@article{CRateHist,
    author = {Samadani, Ehsan and Farhad, Siamak and Panchal, Satyam and Fraser, Roydon and Fowler, Michael},
    year = {2014},
    title = {Modeling and Evaluation of Li-Ion Battery Performance Based on the Electric Vehicle Field Tests},
    volume = {1},
    journal = {SAE Technical Papers}
}

@article{Remus,
    title={Smart Battery Technology for Lifetime Improvement},
    author={Teodorescu, Remus and Sui, Xin and Vilsen, S{\o}ren B and Bharadwaj, Pallavi and Kulkarni, Abhijit and Stroe, Daniel-Ioan},
    journal={Batteries},
    volume={8},
    number={10},
    pages={169},
    year={2022},
    publisher={MDPI}
}

@article{reuter2019importance,
    title={Importance of capacity balancing on the electrochemical performance of {$\text{Li}$}[{$\text{Ni}_{0.8}$}{$\text{Co}_{0.1}$}{$\text{Mn}_{0.1}$}]{$\text{O}_{2}$}{$\,(\text{NCM811})$}/silicon full cells},
    author={Reuter, Florian and Baasner, Anne and Pampel, Jonas and Piwko, Markus and D{\"o}rfler, Susanne and Althues, Holger and Kaskel, Stefan},
    journal={Journal of The Electrochemical Society},
    volume={166},
    number={14},
    pages={A3265},
    year={2019},
    publisher={IOP Publishing}
}

@article{team2008guide,
    title={A guide to understanding battery specifications},
    author={Team, Mit Eletric Vehicle and others},
    journal={Massachusetts Institute of Technology: Cambridge, MA, USA},
    year={2008}
}

@book{groot2012state,
    title={State-of-health estimation of li-ion batteries: Cycle life test methods},
    author={Groot, Jens},
    year={2012},
    publisher={Chalmers Tekniska Hogskola (Sweden)}
}

\end{document}